\documentclass{aa}  

\usepackage{graphicx}
\usepackage{txfonts}
\usepackage{xcolor}
\usepackage{multirow}
\usepackage[skip=9pt plus1pt]{parskip}
\definecolor{azure}{rgb}{0.0, 0.5, 1.0}
\begin{document}

   \title{Dark photon constraints using the UHE gamma-ray emission from galactic sources}

   \subtitle{A Phenomenological  Study}

   \author{A. Melo
          \inst{1},
          S. Hernández \inst{2} \and R. Alfaro \inst{1}
          }

   \institute{Instituto de Física, Universidad Nacional Autónoma de México\\
              \email{arlettemelo@estudiantes.fisica.unam.mx} \and
    Tsung Dao Lee Institute, Shanghai Jiao Tong University \\
             \email{shkdna@sjtu.edu.cn}
             }

   \date{\today}

 
  \abstract
   {Dark photons (Dph) appear in theories beyond the Standard Model of particles (SM). Under certain conditions, it is possible to have a mixing between SM photons and Dphs that should be observed as anomalies in the spectrum of astrophysical sources.}
   {To either find evidence of, or set constraints on the existence of Dphs with masses in the range of $\mu\text{eV}$ using observations of two galactic sources observed at TeV energies.}
   {Using the flux of the Crab Nebula and MGRO J1908+06 at TeV energies reported by HAWC and LHAASO observatories, and assuming a model where Dphs can mix with SM photons in the vacuum; we compute the Test Statistic (TS) to search for evidence of Dphs in the form of variations/attenuation in the observed spectrum.} 
   {We do not find statistically significant evidence of the existence of $\mu\text{eV}$ Dphs. Then, we compute the 68\% C.L. and 95\% C.L. exclusion regions for Dphs with masses in the range from $10^{-8}$ to $10^{-5}~\text{eV}$ and mixing angles with values between 0.01 and 1.0.} 
   {}


   \maketitle
%

\section{Introduction}
\label{sec:intro}

Dark matter (DM) represents almost 27\% of the mass-energy content of the Universe, according to recent observations by \textit{Planck} \citep{Planckcollab}. Although the evidence for the existence of this dark component is well established at several scales of the Universe \citep{bertone2004,ARBEY2021103865,feng_2010}; the nature of DM is not known to this day. Assuming that DM can be explained by an additional particle(s), several candidates have been explored; most of them in the context of extensions to the Standard Model of particles (SM). We can convert the information from astrophysical observations to theoretical constraints in some parameters of the candidates. In particular, the observations require low-velocity particles, also called cold DM (CDM),  to be consistent with the observations. This is included in our modern knowledge of the Universe, the $\Lambda-\text{CDM}$ model. The most studied CDM candidates are massive ($1~\text{GeV}$ to $100~\text{TeV}$) particles whose interactions with SM particles occur at weak scale (Weakly Interactive Massive Particles). However, it has been shown that ultra-light and fuzzy particles ($10^{-20}$ to $10^{-3}$ eV in mass) with integer spin (bosons) can form condensates in the halos around galaxies and behave as CDM particles. Examples of these candidates are Axion-Like Particles (ALPs) and Dark photons (Dphs).

Dphs are hypothetical bosons of a secluded dark/hidden sector and carriers of an interaction similar to the SM-$U(1)$ interaction. Ultra-light ($\mu<10^{-6}\text{eV}$) Dphs have been proposed as CDM candidates to explain entirely (or partially) the DM content in the universe, but heavier ($\mu\thicksim\text{GeV}$) Dphs also appear in the context of multi-component DM scenarios and as mediators of annihilation/decay of heavier DM candidates. For the case of lighter Dphs, a possible communication between dark and SM sectors through portals has been proposed as a possible way to search for evidence of these particles. The communication between both sectors can be modeled as an effective kinetic mixing term in the interaction Lagrangian. The later, allows us to search for possible signals induced by the mixing/oscillations between Dphs and SM photons. Some examples are searches in laboratories using high-intensity lasers in strong magnetic fields \citep{lsw_2007,LSW_milicharged}. Moreover, astrophysical searches for Dphs have been proposed since 1982 \citep{okundphsearches}. 

More recently, conversion probability functions have been derived for several experimental searches. The particular interest are studies from \citep{Sinha}, where the authors derive the conversion probability for Dphs and SM photons in environments with strong magnetic fields, only valid for photons whose energy is close to the mass of the electron. This can be used to search for possible signatures of mixing between Dphs and SM photons in the regions close to pulsars and magnetars, for example. Another interesting result is from the analysis performed by \citep{zechlin}, where they assume that the conversion between Dphs and SM photons can occur in vacuum. They used (at that moment) recent measurments of the flux of the Crab Nebula reported at the GeV--TeV scale. Another astrophysical searches follow the observations of photons at lower energies, for example \citep{2024xray_dph_trident, 2022PhRvR...4a2022X}. 

Here, we aim to update the analysis presented in \citep{zechlin} for the Crab nebula, and show the viability of constraining Dph parameters using the reported fluxes of galactic sources. Our work is different from the one reported by \citep{zechlin}, first of all, because we update the flux measurements for of the Crab Nebula and change the spectral model from Simple Power-Law (SPL) to Log-Parabola (LP) as it better describes the data and the curvature of the spectrum observed at energies above 10 TeV. Secondly, we consider a second source (MGRO J1908+06) with emission reported up to several hundreds of TeV. Furthermore, we estimate the impact of the distance to the source on the exclusion region for MGRO J1908+06. Finally, we compute the expected exclusion regions following MCMC simulations. 

The paper is organized as follows. In Section \ref{sec:dphs}, we briefly discuss Dphs with a particular focus on the conversion probability between Dph and SM photons. We also describe how using observations from astrophysical sources can help us to find bounds to the parameters of Dphs. In Section \ref{sec:sources}, we present the two sources used for the analysis. We also summarize the results of the observations made by HAWC and LHAASO, and the physical parameters relevant to the Dph - SM photon conversions, as the distance to the sources. Then, in Section \ref{sec:modeling}, we discuss the models used as our null and alternative hypotheses. Section \ref{sec:datafitting} describes the methodology used to find the best fit parameters in joint-likelihood analysis, and also how to estimate the impact of the uncertainty on the distance to J1908+063. We present our results and discussions in Sections \ref{sec:results} and \ref{sec:discussion}, respectively. Finally, we present our conclusions in Section \ref{sec:conclusions}.

\section{Dark Photons}
\label{sec:dphs}



Dphs are CDM candidates appearing in theories beyond the SM as particles belonging to a hidden sector additional to the visible. They are the bosons of an electromagnetic-like interaction in the hidden sector, similar to SM photons on the visible sector. While some theories propose a possible interaction between hidden and SM particles; hidden charge is suppressed in the SM sector, and the only possible interaction occurs between Dphs and SM photons.

The communication between both sectors is modulated through a mass state transition (kinetic mixing) between dark and visible photons that  results in the coupling of both particles which occurs during the SM photon propagation through space. Because of this transition between states, it is possible to search for induced effects in the observed spectrum of galactic and extragalactic sources.



Therefore, any attempt to search for these candidates requires knowing the conversion probability between SM photons and Dphs, and what conditions are needed for the mixing to take place. In our case, we use the conversion probability used in light shining through walls experiments, where the weak magnetic field approximation is assumed \citep{lsw_2007}.

\subsection{Photon-Dark Photon coupling}
\label{sub:mixing}

Following \citep{Fabbrichesi_2021}, the Lagrangian to describe the interactions of SM photons and Dphs can be expressed as:
\begin{equation}
     \mathcal{L}_0=-\frac{1}{4} F_{(b)}^{~\mu \nu}  F_{(b)~\mu \nu}-\frac{1}{4} F_{(a)}^{~\mu \nu}  F_{(a)~\mu \nu}-\frac{1}{2} \chi F_{(b)}^{~\mu \nu}  F_{(a)~\mu \nu},
\end{equation}
where $F_{(b)}^{~\mu \nu}$ is the Faraday tensor from the gauge field  $A^{\mu}$ associated to the SM photon, and  $F_{(a)}^{~\mu \nu}$  is the field strength tensor of the electromagnetic-like interaction $U(1)_\text{D}$ on the hidden sector from the gauge field $A^{\mu '}$ associated to the Dph. The term $\chi$ is the kinetic mixing angle that modulates the interaction between both sectors. 



An additional term to the Lagrangian arises from the mass acquisition through a Stueckelberg mechanism \citep{stueckelberg} or the invocation of a ``Dark Higss'' mechanism after the spontaneous breaking of the $U(1)$$_{\text{{D}}}$ symmetry \citep{dark_higgs}. The mass term in the Lagrangian takes the form:

\begin{equation}
    \mathcal{L}_{A^{\mu'}}=\frac{1}{2}\mu^{2} A^{\mu'} A_{\mu}^{'}.
\end{equation} 

After choosing a convenient change of basis and selecting a Lorentz condition for both sectors, the equations of motion matrix representation is the following:

\begin{equation}
\label{eq:motion}
    \bigg[ (\omega^2 + \partial_{z}^{2})   \begin{pmatrix}
1 & 0 \\
0 & 1 
\end{pmatrix} -\tilde{M}  \bigg] \begin{pmatrix}
A \\
A' 
\end{pmatrix} =0,
\end{equation}
where $\tilde{M}$ is a matrix that includes the coupling ($\chi$) and mas ($\mu$) terms given by the following expression:

\begin{equation}
\label{eq:mass_matrix}
     \tilde{M}=\begin{pmatrix}
\chi^2\mu^2& -\chi\mu^2\\
-\chi\mu^2 &  \mu^{2}
\end{pmatrix}.
 \end{equation}

This means that SM photons could convert into Dphs when propagating through space. Please note that in this case we did not use any information about magnetic fields or spatial distribution of charged particles (electrons) in the medium, so this equation describes the propagation of a SM photon-Dph system in the vacuum.

Assuming that we initially have a system of pure SM-photons ($A\neq0, A'=0$) produced at a gamma-ray source, we want to know how many Dphs exists after a time $t$ and traveling a distance $D_\text{L}$. Then, comparing the propagation states matrix of the initial state with the evolved state matrix, the photon-dark photon conversion probability then follows: 
\begin{equation}\label{eq:conversion}
   P_{\gamma\rightarrow\gamma'}(\chi;\mu)=\sin^2(2\chi)\sin^2\bigg(\frac{\mu^2}{4E}D_\text{L}\bigg) \ ,
\end{equation}
where $D_\text{L}$ is the distance from the gamma-ray source to the observation point and $E$ is the energy at which the photon is emitted. The complement of Eq. \ref{eq:conversion} (1-$P_{\gamma\rightarrow\gamma'}$) gives us the number of SM photons after a time $t$ and propagation distance $D_\text{L}$, or simply the survival probability.

\subsection{Indirect searches}
\label{sub:searches}

\begin{figure}[ht]
\centering
\includegraphics[width=9cm]{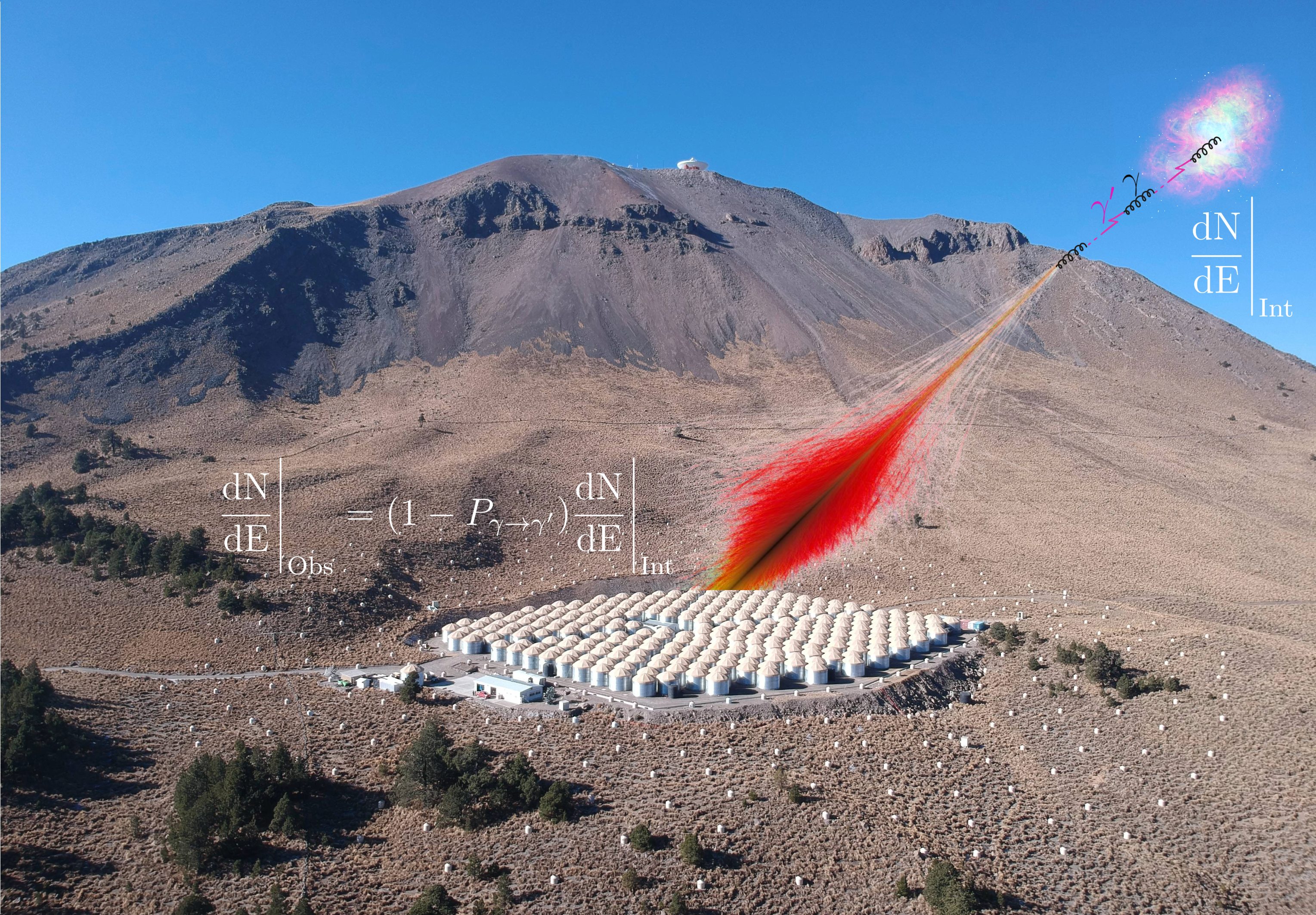}
\caption{Scheme of the proposed search for gamma-ray to Dph oscillation using observations of TeV sources with ground-based gamma-ray observatories. Gamma-rays $\gamma$ produced in a source, as the Crab Nebula, can suffer conversions to DPhs $\gamma'$ in their propagation to Earth. The observed spectrum ($\left.\frac{\rm{d}N}{\rm{d}E}\right|_\text{obs}$) at Earth is related to the intrinsic (produced at the source) spectrum ($\left.\frac{\rm{d}N}{\rm{d}E}\right|_\text{int}$) after including the survival probability of gamma-rays ($1-P_{\gamma\to\gamma'}$).}\label{fig:scheme}
\end{figure}

\begin{figure}[ht]
   \centering
   \includegraphics[width=9cm]{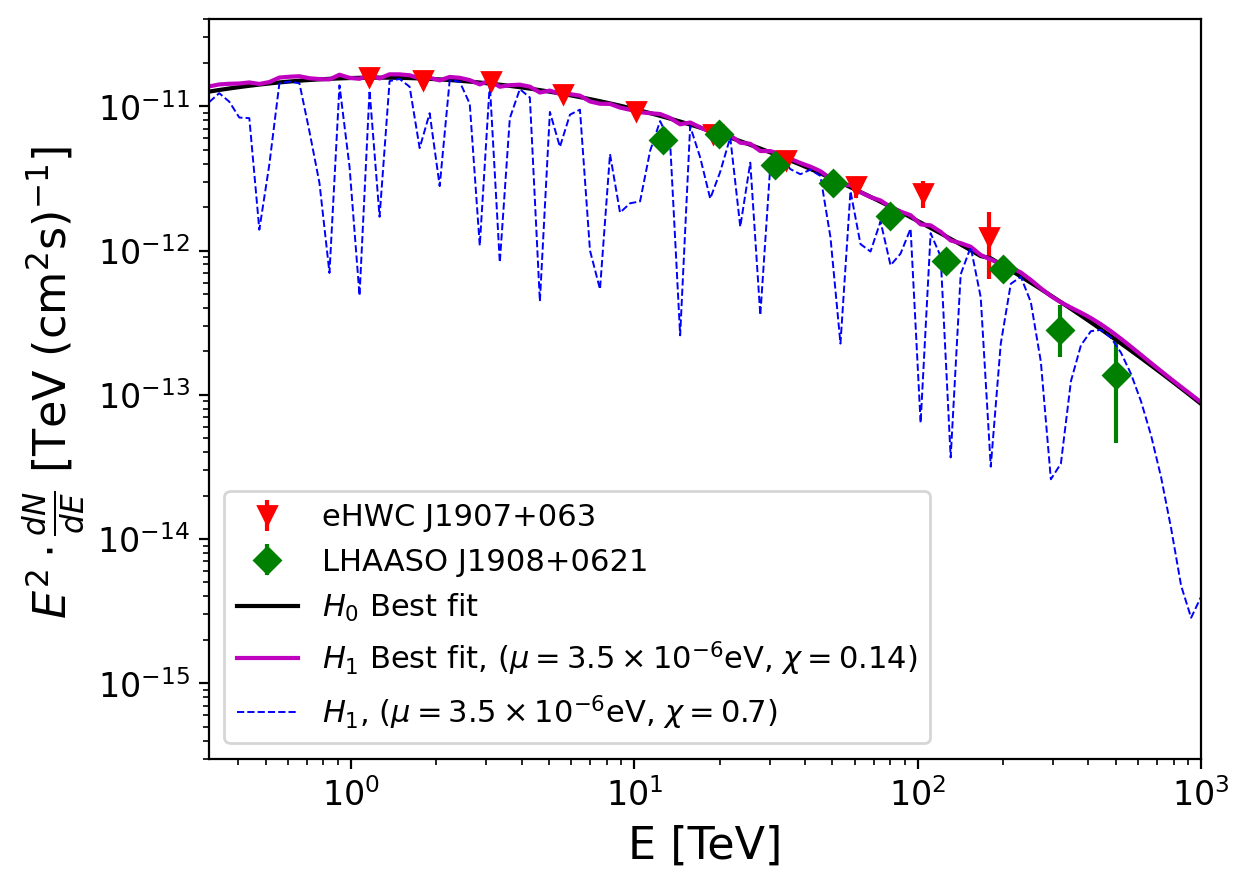}
      \caption{MGRO J1908+06 differential energy spectra plot showing the best $H_0$ fit (black solid line) and the best $H1$  fit (magenta solid line) for the combined data of LHAASO J1908+0621 and eHWC J1908+063, for a dark photon of given a mass of $\mu=1.5\times 10^{-6}$ eV; as well as the fit considering a variation on the dark photon parameters $\chi$ (blue dashed line) to show the possible effect induced by the DM hypothesis on the photon spectra.}\label{fig:coversiones_MGRO}
   \end{figure}

As we mention at the end of the previous section, SM photons, emitted at a gamma-ray source, can undergo oscillations/conversions with Dphs during their propagation through space. Using the survival probability, we can estimate the number of photons per unit of time, per unit of area, per unit of energy (differential flux) that we would observe at Earth. Figure \ref{fig:scheme} shows an schematic view of the proposed search. The effect of SM photon-Dph conversion on the observed flux of MGRO J1908+06 is shown in Figure \ref{fig:coversiones_MGRO}. From Equation \ref{eq:conversion}, we note that the conversion probability depends on the Dph mass $\mu$, the energy $E$ of the SM photon, the mixing angle $\chi$, and the distance $D_\text{L}$ to the source where the SM photons were emitted. If we fix the energy of the SM photon to the energy range of TeV gamma-ray observatories, we can determine the parameter space that we are able to explore by requiring that:

\begin{enumerate}
    \item the distance to the source is large enough to have at least one transition between SM photons and Dphs.
    \item the conversion probability is large enough to observe an effect in the gamma-ray flux.
\end{enumerate}

The first point can be estimated from the oscillation length being directly proportional to $\thicksim E/\mu^2$. The second point depends on the mixing angle and the oscillation length. 


A further requirement to consider a source comes from the fact that SM photons with energies at TeV scales also interact with other radiation fields, such as the Cosmic Microwave Background (CMB), starlight, and Extragalactic Background Light (EBL). The total effect from those interactions is an extra attenuation to the observed flux. In the end, we need to model this secondary opacity of TeV photons related to the different radiation fields involved, according to the luminic distance $D_\text{L}$ of the gamma-ray source. Each radiation field will introduce additional uncertainties in the analysis, and our goal is to reduce them to the minimum. Then, we consider only galactic TeV sources. This means that only photons with energies above hundreds of TeV suffer from attenuation of radiation fields \citep{2021Natur.594...33C}.

In agreement with the fixed energy of the SM photons and the distance to the sources ($\thicksim\text{kpc}$), the parameter space of Dphs is limited to the range of masses $\mu$ from $10^{-8}~\text{eV}$ to $10^{-5}~\text{eV}$, and mixing angles $\chi$ larger than 0.001.


\section{Astrophysical Sources}
\label{sec:sources}


Based on the criteria described in Section \ref{sub:searches}, we chose two galactic sources with emission at TeV scales observed by current gamma-ray observatories. In this section, we present the observations for these sources and the relevant parameters to our study.

\subsection{Crab Nebula}
\label{sub:crab}

The Crab is a system located in the Taurus constellation at a distance of $2$kpc from Earth, resulting from a supernova event. It consists of the Crab pulsar, a neutron star with strong magnetic fields and high velocity rotation; the Crab Nebula,  a pulsar wind nebula result of the interaction of the electrons ejected from the pulsar (pulsar wind) with the medium; and the remnant of the explosion \citep{crab_review}. 

The Crab Nebula's nearly constant emission has been observed throughout the whole range of the electromagnetic spectrum, widely contributing to a deeper understanding of the processes involved in the emission of high energy photons that reach the PeV energy scale.
The main processes involved in the nebula's very-high energy (VHE) ($E\sim 10\ \text{TeV}$)  emission above $60$ GeV are Synchrotron and Inverse Compton, as well as Synchrotron Self-Copton \citep{sscompton}. The two later are mainly associated with its ultra-high energy (UHE) ($E\sim 100 \ \text{TeV} - 1\ \text{PeV}$)  radiation.   
 
These unique characteristics have earned it the title of standard candle for the calibration and design of all telescopes and cosmic ray observatories currently operating. Both, the Crab Pulsar and the Crab Nebula, have been subject to a great amount of studies in the cosmic rays and high-energy astrophysics field in general ever since its first observation in AD 1054.

For this source, we are using the latest observations from HAWC \citep{2019ApJ...881..134A} and LHAASO KM2A and WCDA \citep{2021ChPhC..45h5002A, 2024ChPhC..48f5001C} of the spectral energy distribution to obtain a dataset consisting of energies from $500$ GeV and including LHAASO's PeV photon. The Crab Nebula´s spectral dataset is usually described by a log-parabola model, with the observatories' individual estimated parameters $\phi_0=(2.35\pm 0.04) \times 10^{-12}\text{cm}^2 \text{TeV}^{-1} \text{s}^{-1}$, $\alpha=2.79\pm 0.02$, $\beta=0.10\pm 0.01$ found by HAWC \citep{2019ApJ...881..134A} using a ground-parameter estimator, and with LHAASO's estimated parameters $\phi_0=(8.72 \pm 0.10) \times 10^{-12}\text{cm}^2 \text{TeV}^{-1} \text{s}^{-1}$, $\alpha=2.92 \pm 0.04$, $\beta=0.18 \pm 0.04$ for the KM2A array \citep{2024ChPhC..48f5001C}, and $\phi_0=(2.32\pm 0.19) \times 10^{-12}\text{cm}^2 \text{TeV}^{-1} \text{s}^{-1} $, $\alpha=2.57\pm 0.06$, $\beta=0.02\pm 0.05$ for the WCDA array \citep{2021ChPhC..45h5002A}.

\subsection{MGRO J1908+06}
\label{sub:j1908}


Another interesting extended UHE gamma-ray source,  that would offer a complementary study due to its specific characteristics, is MGRO J1908+06. This is a source seen both by LHAASO and by HAWC, reported as LHAASO J1908+0621 in \citep{2021Natur.594...33C} and eHWC J1908+063 in \citep{PhysRevLett.124.021102}. It was first seen by Milagro, but its emission has also been detected by other gamma-ray observatories like HESS, ARGO, VERITAS, HAWC and LHAASO. Its emission extends past the PeV scale. However, there is not a clear association to this source. Several distances (from Earth to source) have been proposed according to possible counterparts such as:  $2.37$ kpc  \citep{Manchester_2005} linked to radio-faint pulsar PSR J1907+0602, $3.4$ kpc associated to the center of radio supernova remnant SNR G40.5-0.5, and $8.8$ kpc the distance to pulsar PSR J1905+0600 \citep{hawcJ1908}.


Whether the nature of the source remains unknown, for this study, we decided to use the distance reported to SNR G40.5-0.5 as a benchmark model. In addition, we explore the impact of the change in the distance to the source on the final Dph constraints.

In this study, we are contemplating MGRO J1908+06 spectral energy distribution observations reported by both HAWC \citep{PhysRevLett.124.021102} and  LHAASO (only KM2A) \citep{2021Natur.594...33C}, to build a second dataset with the data's energy range going from $1$ TeV and up to $500$ TeV. This source's spectra is usually described by a log-parabola model as well, with HAWC's estimated parameters $\phi_0=(0.95 \pm 0.05) \times 10^{-13}\text{cm}^2 \text{TeV}^{-1} \text{s}^{-1} $, $\alpha=2.46\pm 0.03$, $\beta=0.11\pm 0.02$, and LHAASO's $\alpha=2.27$, $\beta=0.46$ estimated parameters.


\section{Spectral Models}
\label{sec:modeling}

For the analysis, we use two spectral models to describe the data and search for evidence of the existence of dark photons. Each spectral model corresponds to a different hypothesis to test: a null hypothesis $H_0$, where we do not assume existence of Dph; and an alternative hypothesis $H_1$, where we consider SM photon-Dph oscillations. In the following, we describe the parameters of each model.

\subsection{Null Hypothesis $H_0$}

For the null hypothesis, we refer to the case where the gamma-ray flux is well described by the standard astrophysical processes occurring at the position of the source (acceleration, diffusion, and others). We can refer to this model as the intrinsic spectrum. Moreover, we should consider other propagation effects, as $\gamma-\gamma$ interactions, that can lead to attenuation of the TeV photons emitted at the source. However, we do not know all the processes in order to extensively explain the observed gamma-ray flux. In order to avoid a model-dependent description of the observed flux, we use a phenomelogical approach, where the observed flux is described according to the best model obtained after fitting the observations. In particular, the spectra of the Crab Nebula and MGRO J1908+06 are well described by a log-parabola model (see Section \ref{sec:sources}) for energies above $\sim\text{GeV}$ as it properly matches the VHE and UHE gamma-ray emissions. The log-parabola follows the equation:


\begin{equation}\label{eq:log-parabola}
    \frac{dN}{dE}=\phi_0 \bigg(\frac{E}{E'}\bigg)^{-\alpha-\beta \ln\big(\frac{E}{E'}\big)},
\end{equation}
where $\phi_0$ is the normalization of the spectra, $E'$ is the pivot energy, $E$ is the energy of the gamma-ray, $\alpha$ is the spectral index and $\beta$ is the curvature index.

To consider the attenuation effects due to the propagation of TeV gamma-rays in our galaxy, we consider the attenuation, as a function of the energy, used in the analysis by the LHAASO collaboration of the Crab Nebula and MGRO J1908+06 \citep{2021Natur.594...33C}.

The complete model used for the null hypothesis is:

\begin{equation}\label{eq:full_null_model}
    \frac{dN}{dE}=\phi_0 \bigg(\frac{E}{E'}\bigg)^{-\alpha-\beta \ln\big(\frac{E}{E'}\big)}\times f_\text{att.}(E),
\end{equation}
with $f_\text{att.}(E)$ encoding the attenuation of TeV gamma-rays induced by $\gamma-\gamma$ interactions.

\subsection{Alternative Hypothesis $H_1$}\label{subsec:alternative}

Additional to the log-parabola and the $\gamma-\gamma$ interaction considered for the null hypothesis, the alternative hypothesis includes the survival probability of TeV gamma-rays obtained from the conversions between Dphs and SM photons when propagating towards Earth. In this case, the model is described by:

\begin{equation}\label{eq:full_alternative}
    \frac{dN}{dE}=\Big(1-P_{\gamma\rightarrow\gamma'}(\chi;\mu)\Big)\times\Phi_0 \bigg(\frac{E}{E'}\bigg)^{-\alpha-\beta \ln\big(\frac{E}{E'}\big)}\times f_\text{att.}(E),
\end{equation}
where $P_{\gamma\rightarrow\gamma'}$ follows from Equation \ref{eq:conversion}.

\section{Model Fitting Methodology}
\label{sec:datafitting}

In this section, we describe the different steps to find the best fit parameters for the null and alternative hypotheses and to obtain the exclusion upper limits (U.L.).




\subsection{Maximum Likelihood Estimates}

The maximum likelihood (ML) method allows us to determine the probability that a model of interest $\lambda$, with parameters $\Vec{\theta}$, explains an experimental dataset. The likelihood function $\mathcal{L}$ is the product of the normalized probability density function at the given measurements $x_i$ with uncertainties $\sigma_i$:

\begin{equation}\label{eq:likelihood}
\mathcal{L}(\Vec{\theta};\{x_i,\sigma_i\}) = \prod^n_{i=1} p(x_i;\Vec{\theta},\sigma_i^2),
\end{equation}

Optimization of $\mathcal{L}(\Vec{\theta};\{x_i,\sigma_i\})$ provides the estimates of the model parameters that maximize the probability of the model describing the data, so that $\lambda (\hat{\Vec{\theta}})$ is the value of the model evaluated with the best-fit parameter values.


We consider the fluxes and their statistical uncertainties reported by the HAWC and LHAASO collaborations for the Crab Nebula and MGRO J1908+06.  Then, we use a Gaussian distribution to estimate the likelihood given the observed fluxes. The log-likelihood for one dataset is:

\begin{equation}\label{eq:log_likelihood}
\ln\left(\mathcal{L}(\Vec{\theta};\{x_i,\sigma_i\})\right)= -\frac{1}{2}\sum_{i=1}^n\Bigg(\frac{(x_i-\lambda(\Vec{\theta}))^2}{\sigma_i^2}+\ln(2\pi\sigma_i^2)\Bigg),
\end{equation}
where $\lambda(\Vec{\theta})$ is the model expectation of the flux, and ${x_i}$ and ${\sigma_i}$ are the observed fluxes and statistical uncertainties, respectively. The log-likelihood is summed over the $n$ energy intervals.


As the observations from different experiments are assumed to be independent between each other, the Joint Likelihood (JL) function takes the product of the individual likelihood functions. Using the logarithmic of the functions, this translates into a sum over the different datasets:
\begin{equation}\label{eq:joint_likelihood}
\ln\left(\mathcal{L}_\text{\tiny{joint}}\right)=\sum_{\text{\tiny exp}}\ln\mathcal{L}_\text{\tiny{exp}}(\Vec{\theta}).
\end{equation}

For the same source, we assume the model $\lambda$ used to explain the reported flux by HAWC to be the same model to explain the LHAASO's flux points. Then, both datasets share the same model parameters $\Vec{\theta}$.

\subsection{Test Statistic}\label{sec:ts}


We use the Likelihood Ratio Test (LRT) and the Test Statistic ($TS$), to perform a comparison between our null and alternative hypothesis and decide which model provides a better description of the data. The $TS$ as a function of the mixing angle $\chi$, for a given Dph mass $\mu$, is:


\begin{equation}\label{eq:ts}
    TS(\chi;\mu)=-2\ln\left(\frac{\mathcal{L}_{0}(\Vec{\alpha,\beta,\phi_0)}}{\mathcal{L}_{1}(\Vec{\alpha',\beta',\phi_0'},\chi;\mu)}\right).
\end{equation}
The parameters $\Vec{\alpha,\beta,\phi_0}$ are the best-fit parameters for the null hypothesis, while the parameters $\Vec{\alpha',\beta',\phi_0'}$ are the best-fit parameters obtained for the alternative hypothesis as a function of the mixing angle $\chi$, for a fixed $\mu$.

\subsection{Data selection and JL fit for the null hypothesis}\label{sec:dataselection}
When performing a joint analysis with observed data from different observatories, we must have in mind that each observatory has different characteristics and sensitivities. Thus, the different datasets require a supplementary preparation state based on the individual detector response functions. We perform a scaling of the HAWC and LHAASO data to take into account the differences of each observatory as we do not have direct access to the counts maps. The scaling factors are obtained by a JL fit. We only consider the flux reported by the HAWC and LHAASO collaborations because we want to reduce the systematic uncertainties, and the correlation between the different parameters of the null hypothesis\footnote{We also considered reported fluxes from HESS, MAGIC, Tibet-$\gamma$AS, and Veritas. However, the uncertainties of the parameters for the null hypothesis and the correlation between normalization $\phi_0$ and spectral index $\alpha$ were minimal in the case of only considering HAWC and LHAASO measurements}.



After selecting an adequate dataset, we repeat the fit for the null hypothesis over a list of 100 different fixed values for the pivot energy $E'$,  in the energy range from $1$ TeV to $100$ TeV, to find the value that minimizes the correlation between the spectral index $\alpha$ and the normalization $\phi_0$. Once we have the best scaling factors and pivot energy $E'$, we repeat the JL fit to find the best-fit parameters for the null hypothesis.

\subsection{JL fit for the alternative hypothesis}
\label{fit_althyp}

For the alternative hypothesis, we consider a mesh of DPh masses $\mu$ in the range from $10^{-8}~\text{eV}$ to $10^{-5}~\text{eV}$. For each value, we perform a JL fit to find the best-fit parameters given our model described in Equation \ref{eq:full_alternative}. Furthermore, we compute the likelihood profiles (Equation \ref{eq:ts}) for mixing angles $\chi$ in the range from 0.001 to 1.0. 

In the case where we do not find any evidence of the existence of DPhs ($TS<3$), and assuming that likelihood profiles are parabolic, we can define $\Delta TS$ as:

\begin{equation}\label{eq:deltats}
    \Delta TS = TS_{\text{max}}-TS(\chi;\mu),
\end{equation}
with $TS_\text{max}$ the value of the test statistic evaluated at the best fit-parameters of the null hypothesis and the alternative hypothesis (for fixed $\mu$). For the range of parameters we are exploring, in most of the cases, the $\Delta TS$ follows a $\chi^2$ distribution with 1~\textit{d.o.f.}. We use the $\Delta TS$ to find the U.L. (95\% C.L. and 68\% C.L.) of the mixing angle $\chi$ for each mass $\mu$ tested.

\subsection{Expected limits}
\label{sec:expected_limits}

To estimate the impact on the exclusion region induced by statistical fluctuations in the photon spectrum (related to characteristics of the observatories, miss-classification of cosmic ray events, etc.); we generate a sample of 200 simulated spectra of the Crab Nebula. We fluctuate the number of photons, as a function of the energy, assuming a Gaussian distribution, for each reported data set (HAWC, LHAASO), and apply the same fit methodology reported in this section: scaling factor, joint likelihood for null and alternative hypotheses, calculation of TS, and get the exclusion regions (95\% C.L. and 68\% C.L.) on the dark-photon mixing angle $\chi$. For the resultant distribution of the expected U.L. we estimate the mean value, $1\sigma$ and $2\sigma$ confidence intervals (C.I.). We apply an additional bootstrapping of the resulting U.L. distributions to smooth the final values of the expected exclusion regions.


\section{Results}
\label{sec:results}

In this section, we present the different results for the null and alternative hypotheses.  The underlying fits were performed using the \texttt{iMinuit} $2.25.2$ minimizer \citep{iminuit}\footnote{https://github.com/iminuit/iminuit}.

\subsection{Null hypothesis fits}
\label{sub:null_and_alt}

Previous to performing the hypotheses fits, the scaling factors presented in Table \ref{tab:scaling} were found for each observatory, for each dataset. The scaling factors obtained for LHASSO's observations are close to one.  

\begin{table}[ht]
\centering{
\begin{tabular}{ |p{2cm}|p{2.7cm}|p{2.7cm}| }
\hline
 & $s$  & $s$ \\
 
  Observatory &(Crab Nebula)& (MGRO J1908+06)\\
 \hline 
 
HAWC & 1.00  & 1.00\\
LHAASO &  1.02 &0.86\\
\hline
\end{tabular}
\caption{Scaling factors obtained from the fit taking HAWC data as a reference.}
\label{tab:scaling}
}
\end{table}

\begin{table*}[t]
\centering
\begin{tabular}{|l|l|l|l|l|l|l|l|}
\hline
\multirow{2}{*}{Source} & \multirow{2}{*}{\begin{tabular}[c]{@{}l@{}}Distance\\  {[}kpc{]}\end{tabular}} & \multirow{2}{*}{RA {[}$^{\circ}${]}} & \multirow{2}{*}{Dec {[}$^{\circ}${]}} & \multirow{2}{*}{\begin{tabular}[c]{@{}l@{}}$E'$\\  {[}TeV{]}\end{tabular}} & \multirow{2}{*}{\begin{tabular}[c]{@{}l@{}}log $\phi_0$\\ {[}(TeV cm$^2$s)$^{-1}${]}\end{tabular}} & \multirow{2}{*}{$\alpha$} & \multirow{2}{*}{$\beta$} \\
                        &                                                                                &                                      &                                       &                                 &                                                                                                    &                           &                          \\ \hline
Crab Nebula             & $\sim 2.0$                                                                     & $83.61 \pm 0.02$                     & $22.00 \pm 0.03$                      & $4.05$                          & $-11.982 \pm 0.013$                                                                                & $2.696 \pm 0.020$          & $0.097 \pm 0.010$         \\ \hline
MGRO J1908+06          & $3.4$ to $8.8$                                                            & $286.91 \pm 0.10$                    & $6.32 \pm 0.09$                       & $6.43$                          &     $-12.553 \pm 0.015$                                                                                               &   $2.381 \pm 0.028$                        &    $0.117 \pm 0.012$                      \\ \hline
\end{tabular}
\caption{The spectral fit values for the Crab Nebula and for MGRO J1908+06, obtained from the Log-parabola MLE, are presented. Additionally, the right ascension (RA) and declination (Dec) information for each source is provided.}
\label{tab:null_hyp}
\end{table*}

For each source, the estimated parameters of the joint differential energy distribution, without considering a photon-dark photon conversion, are reported in Table \ref{tab:null_hyp}. The position parameters of the source are those reported by the observatory's analysis, and the pivot energies ($E'$) are those found from minimizing the correlation between normalization and the spectral index $\alpha$. For the normalization $\phi$, the spectral index $\alpha$, and the curvature index $\beta$, the central values and their statistical uncertainties ($1\sigma$) are those from the best fit.


\subsection{Alternative hypothesis: exclusion regions}

Figure 2 illustrates an example of the fits performed for the alternative hypothesis for a specific mass value ($\mu=3.5\times 10^{-6}$ eV) and its comparison to the null hypothesis. For the alternative hypothesis, when considering a mixing angle value of $\chi=0.7$ (dashed blue) we would expect quantitatively significant oscillations, even at PeV energies, that are not seen by the observatories. However, for the $H_1$ best fit parameters (solid magenta), these oscillations are not so evident and are actually comparable to the $H_0$ best fit (solid black).

\begin{figure}[ht]
   \centering
   \includegraphics[width=9cm]{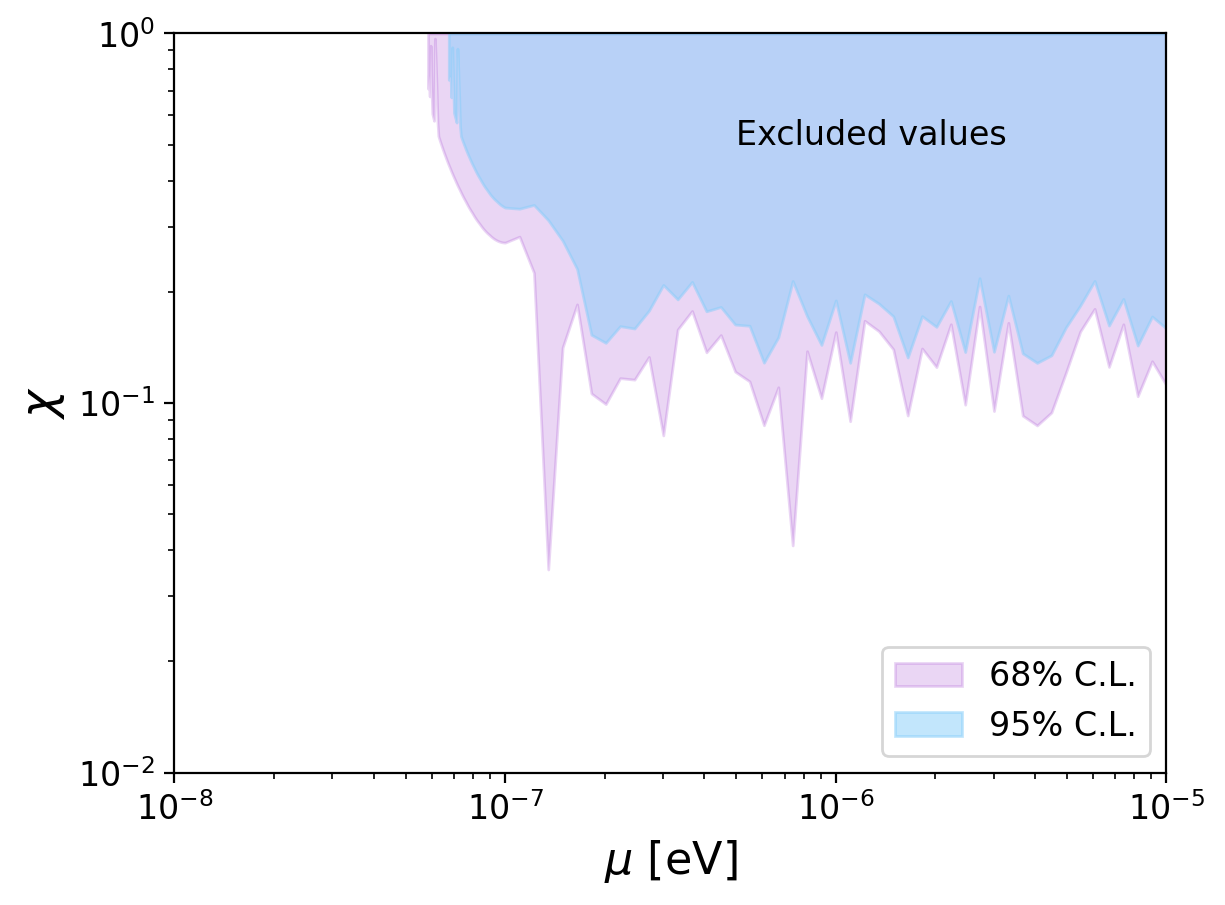}
      \caption{Exclusion region of the parameter space of dark photons at a 95\% C.L. (blue) and 68\% C.L. (pink) using the VHE gamma-ray emission of the Crab Nebula. }\label{fig:exclusion_crab}
   \end{figure}

\begin{figure}[ht]
   \centering
   \includegraphics[width=9cm]{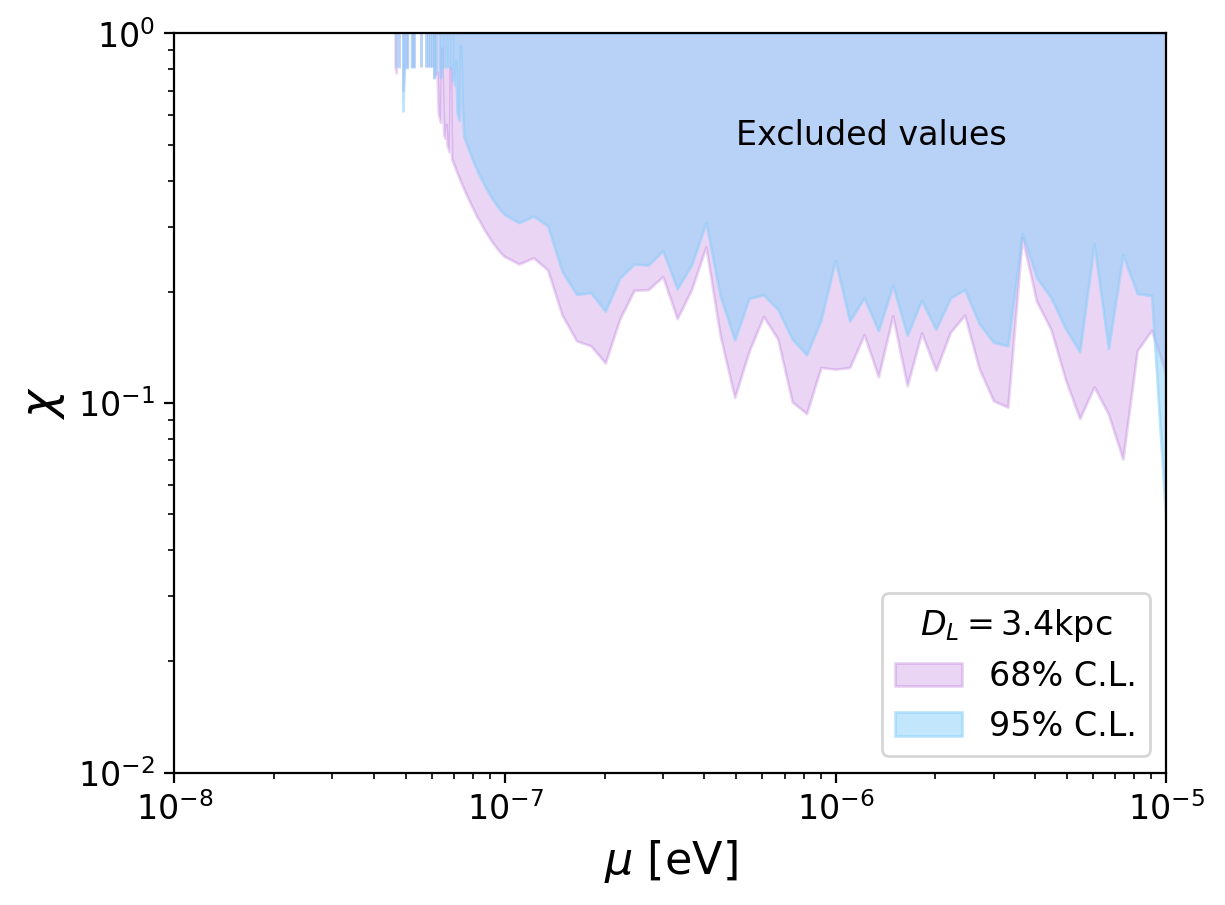}
   
      \caption{Exclusion region of the parameter space of dark photons at a 95\% C.L. (blue) and 68\% C.L. (pink) using the reported gamma-ray spectra of MGRO J1908+06 considering a luminic distance of $D_L=3.4\ \text{kpc}$.}\label{fig:exclusion_j1908}
   \end{figure}

From Figure \ref{fig:exclusion_crab} and Figure \ref{fig:exclusion_j1908} the parameter space exclusion regions of both sources exhibit a similar trend with scarce information on the mixing angle limits for masses below $7.7 \times 10^{-8}$ eV and oscillating U.L. values throughout the entire range of the exclusion region.

\begin{figure}[ht]
   \centering
  \includegraphics[width=9cm]{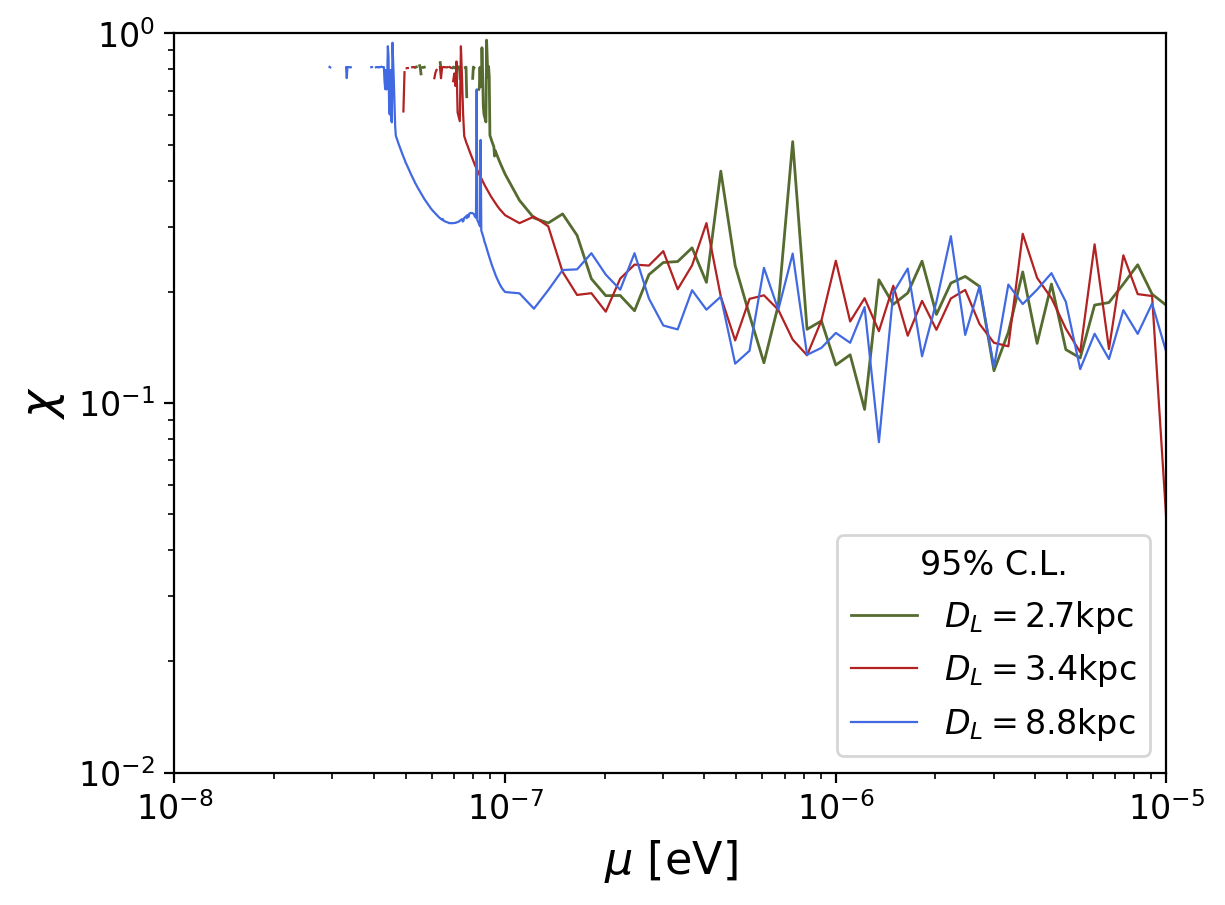}
      \caption{Exclusion region of the parameter space of dark photons at a 95\% C.L. for different luminic distance values.}\label{fig:exclusion_j1908_dist}
   \end{figure}

    In the same way that the analysis was done for each source and for a fixed luminic distance, an analysis on MGRO J1908+06 for 3 different lumininc distances was replicated. Figure \ref{fig:exclusion_j1908_dist} illustrates how variations in the distance to the source can alter the final constraints of the Dph parameters. In particular, for the case of $D_L=8.8$ kpc we see that the exclusion region extends to lower masses ($4.02 \times 10 ^{-8}$ eV).

\begin{figure}[ht]
\centering
    \includegraphics[width=9cm]{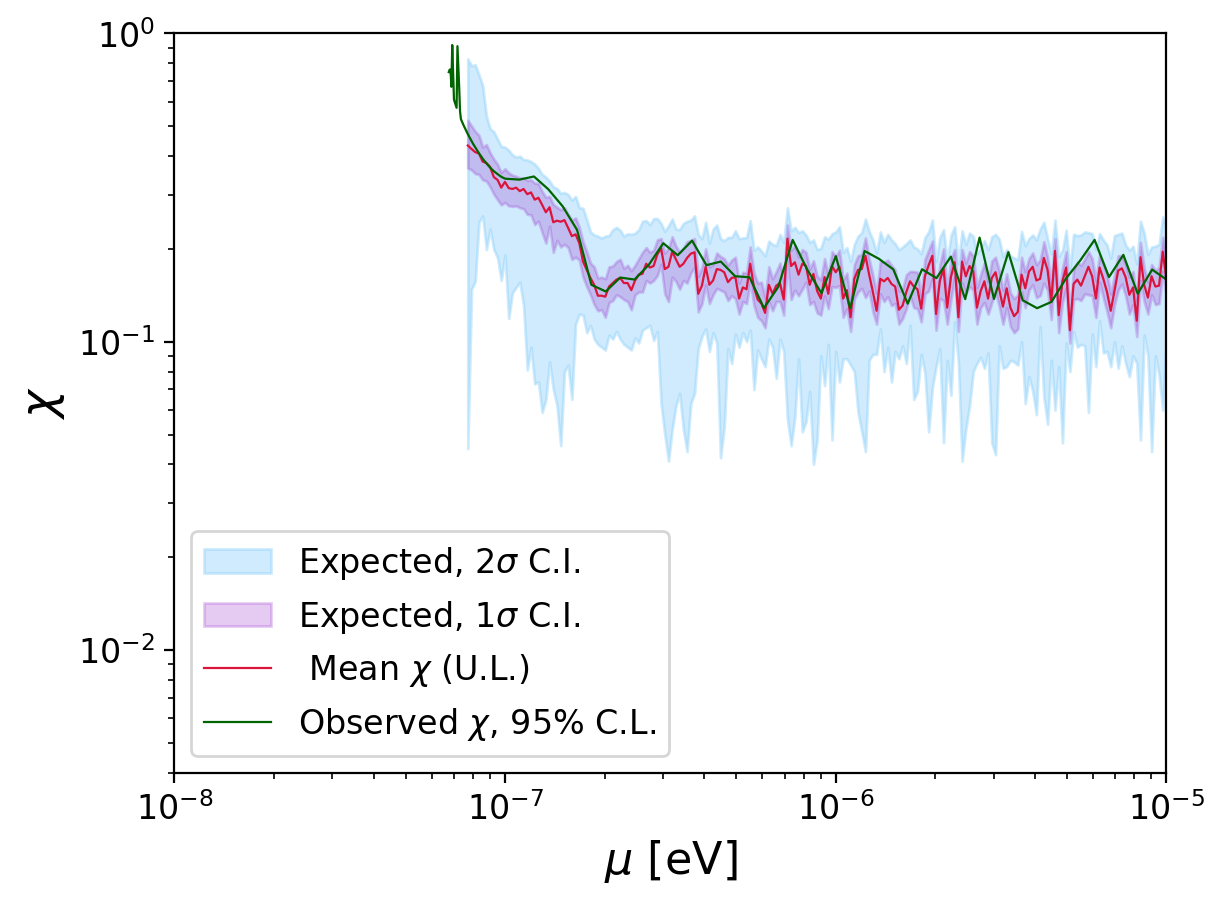}
    \caption{Crab Nebula's expected region for the excluded mixing angle and Dph mass paired values.}
    \label{fig:expected_CRAB}
\end{figure}

 In Figure \ref{fig:expected_CRAB}, we present the resulting expected regions . The expected region for a $2 \sigma$ C.I. fully contains the  $1 \sigma$ C.I. region  and is consistent with the U.L. from the observations. In addition, the mean value of the simulated spectra U.L. mixing angles do not differ significantly from the U.L. values derived directly from the reported data analysis.


\section{Discussion}
\label{sec:discussion}

\subsection{Null hypothesis}
The $H_0$ fit performed using ML uses the same spectral hypothesis with a change in the pivot energy from those used by the observatories. In result, we get mismatched normalizations. Regardless of using mismatched pivot energies between the ML fit and the fits performed by the observatories, resulting in unalike normalization values; the spectral hypothesis used in both cases remains the same and the spectral and curvature indices stay consistent with HAWC's and LHAASO's measurements. 

From \cite{2021Natur.594...33C}, it is shown that the magnitude of the attenuation due to the cosmic microwave background (CMB) and interstellar radiation field (ISRF) is actually comparable to the magnitude of the attenuation presented as a product of photon-dark-photon conversions. However, in this study we have found that after including the corresponding attenuation factor $f_\text{att}(E)$ in the spectral model from Eq. \ref{eq:full_alternative}, there was no significant difference from the final estimated values previously found for both $H_0$ and $H_1$ spectral fit parameters.

\subsection{Alternative hypothesis and expected limits}

 The obtained exclusion regions (at 68\% C.L.), for both Crab Nebula and MGRO 19086+06, have a similar morphology to the constrains reported by \citep{zechlin}; with a notorious difference in the extension of the regions. In \citep{zechlin}, it appears to report a more restrictive result. However, for this study we are actually contemplating an updated dataset with energies over $10$ TeV. This means that the model fitted has to be adequate for VHE emissions, changing from a simple power-law to a log-parabola spectral hypothesis. As a result, the exclusion regions in this article have a shift in the range that we are able to explore to larger mixing angles and larger masses. In particular, for the specific mass range from $10^{-8}$ to $10^{-7}$ eV, we observe a discontinuity on the search for UL values. This is largely due to the $\Delta TS$ profile not being adequate (parabolic) and/or not reaching the $\Delta TS$ criteria to calculate the 95\% and 68\% U.L.

The excluded regions we found from the analysis for the Crab Nebula and MGRO J1908+06 correspond to a region of the parameter space where the DPh does not explain all the DM in the Universe, see Fig. 4 on \citep{PaolaArias_2012}. However, models where the DPh does not explain the DM relic density are still viable, corresponding to multi-component scenarios. Examples are models with a massive fermion DM particle, where Dph can be produced through annihilation \citep{multidphcomponent}.



As for the expected regions, performing a bootstrapping procedure on the Crab Nebula's spectral iterations is comparable to increasing the number of simulated spectra, as this serves the intended purpose of softening the U.L. distribution while also reducing the computation time and enhancing the computing in general. What these expected regions ($1\sigma$ and $2\sigma$ C.I.) depict is that the observed constraints are in agreement with changes within the statistical uncertainties of the observed spectra in the way that we would not be able to distinguish effects induced by the mixing between gamma-rays and DPh candidates inside the expected exclusion region.

\section{Conclusions}
\label{sec:conclusions}

   Using the gamma-ray emission at TeV and up to PeV energies of two different sources we were able to establish bounds to the Dph parameters. We used the ML method to estimate the best fit parameters of two different spectral hypotheses: a model independent depiction of the emitted photons and a model including the effect induced by Dphs. By comparing both hypotheses and replicating the analysis for simulated spectra, we obtained exclusion regions that rule out Dph candidates on the mass range from $10^{-8}$ to $10^{-5}~\text{eV}$ and mixing angle between 0.01 and 1.0.
   
   This allows us to consider Dph searches in sources with similar characteristics and/or doing a joint analysis with data from the counts maps of multiple experiments.  Another interesting approach to studying the effects of photon-Dph conversions would be to focus on a model dependent analysis that provides a physical description of the intrinsic spectra  instead of assuming a log-parabola spectral hypothesis to explain the spectral energy distribution (SED) of the Crab Nebula, as proposed by  \citep{2023A&A...671A..67D}. Additionally, other model-independent searches based on irregularity tests of the gamma-ray SED \citep{irregularitytest} should also offer complementary results to methods based on model-comparison tests, as the one we have used on this work.

    For this study we assume that gamma-rays travel in an empty space, without the presence of magnetic fields and charged particles in the interstellar medium. While the approximation for null magnetic field seems to be valid for TeV photons, the conversion probability obtained in \citep{lsw_2007} for non-zero magnetic field may not be valid at TeV energies. More studies are needed to properly estimate the conversions between TeV gamma-rays and DPhs in the interstellar medium.

    In addition to the previous considerations, we should be able to improve the constraints on the mixing angle $\chi$ by using the observations of extragalactic sources (Active Galactic Nuclei or Gamma-ray burst). In particular, it is of great importance to reach the region where the DPh could account for all the DM in the Universe. This also implies to proper modeling the different magnetic fields during the propagation of gamma-rays. This is similar to the existing framework dedicated to search for evidence of ALPs, and we may adapt/integrate those methods to search for DPhs at TeV scales.

\begin{acknowledgements}
      Part of this work was supported by DGAPA-UNAM through the 
      \emph{ Programa de Apoyo a Proyectos de Investigación e Innovación Tecnológica (\textbf{PAPIIT})} program, project AG101323, and NSFC Grant No.12393853.
\end{acknowledgements}

\bibliographystyle{aa} 
\bibliography{mybib} 

%
%






   
  



\end{document}